\documentclass[a4paper,twoside]{article} 
\input{vssc.sty} 

\usepackage{multirow}

\begin{document}
	
\VSSCtitle{NSV 4031 is a gamma Doradus variable after all}

\VSSCauth{
	Christopher Lloyd\\	
	cl57@ymail.com
}
	

%
%
%
\VSSCabstract{NSV 4031 has already been comprehensively debunked as an eclipsing binary, but the data available at that time only allowed a relatively poor limit of a few hundredths of a magnitude on any other variation. Analysis of more recent TESS data shows that the star is a multiperiodic variable, with dominant periods in the 1-day range and semi-amplitudes of 3 mmag. In total some 19 periods have been recovered with semi-amplitudes above 0.1 mmag, suggesting that the star is a gamma Doradus variable,  with no detectable variations at frequencies above 2 c/d.}

\section*{Introduction}

\href{https://simbad.cds.unistra.fr/simbad/sim-id?Ident=NSV+4031}{NSV 4031 (HD 70271, BD +45\degr1570, CSV 6628, SVS 1396 Lyn)} is a ninth magnitude star in Lynx and was reported as a probable Algol-type eclipsing binary by
\citet{1963ATsir.246....3L}. 
The report is based on a few days of visual observations made during dusk, and claimed a variation of 0\fmm8 on a likely period of about half a day. As a variable star NSV~4031 has been largely ignored, but it was included in the BAA VSS Eclipsing Binary Programme, in an effort to clarify its behaviour. Limited visual observations during 2003 showed no significant variation. In a detailed review of all the available data from the archives of the synoptic survey projects Tycho, ROTSE-I (NSVS), SuperWASP, ASAS-SN, TASS and the INTEGRAL OMC, plus new, dedicated observations, 
\citet{2020OEJV..207....1L} showed there was no evidence that the star was an eclipsing binary as suggested, or indeed that there was any variation above a few hundredths of a magnitude. The star is now given as 'constant' by \href{https://vsx.aavso.org/index.php?view=detail.top&oid=42655}{VSX}, but is still listed as an eclipsing binary by Simbad.

\section*{TESS data}

Since the previous paper, data from the Transiting Exoplanet Survey Satellite (TESS, \citet{2015JATIS...1a4003R}) have become available for NSV~4031, which was observed in sectors 20, 47 and 60, in January 2020, 2022 and 2023, at the 30-minute, 10-minute and 200-second cadence respectively. 
The data were downloaded from the
\href{https://mast.stsci.edu/portal/Mashup/Clients/Mast/Portal.html}{MAST archive at the STScI}, and both the 
high level science products (HLSP) Quick Look Pipeline (QLP) and
\href{https://heasarc.gsfc.nasa.gov/docs/tess/pipeline.html}{HLSP TESS-SPOC} photometry pipeline data have been examined.
Each sector is observed over two 13.7-day orbits, but with, in this case, a few days lost each orbit for data download. Other small gaps can appear at the half-orbit points due to the high background, and limitations in the background removal frequently generates systematic trends or offsets in the fluxes between the two orbits in each sector, and even in the halves of each orbit.

\begin{figure}[!b]
	\centering
	\includegraphics[width=0.49\textwidth]{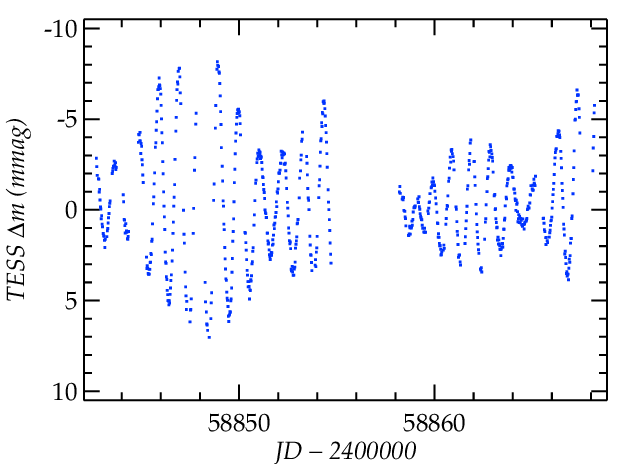}
	\includegraphics[width=0.49\textwidth]{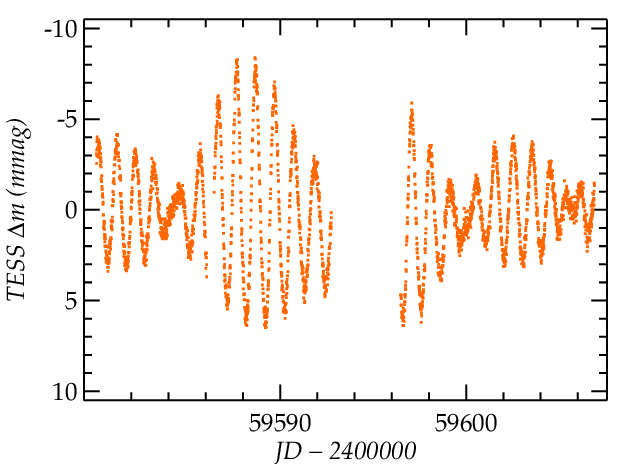}
	\includegraphics[width=0.49\textwidth]{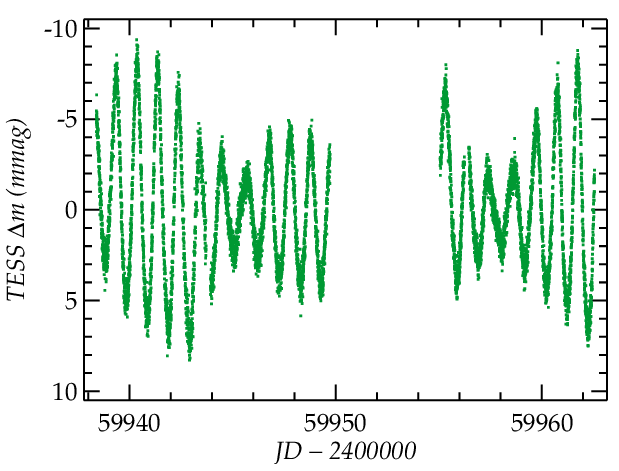}
	\hfill
	\caption{The TESS QLP-SAP light curves of NSV 4031 for sectors 20, 47 and 60, showing that the star is clearly a low-amplitude multiperiodic variable. The HLSP light curves are very similar, but with significant gradients through the different orbits, and the variation is largely suppressed in the QLP-KSPSAP photometry. \label{fig:epoch}}
\end{figure}

Both the QLP and SPOC pipelines generate two variants of the target flux based on different approaches to flux extraction and background correction. These are the raw, simple aperture photometry, QLP-SAP and SPOC-SAP fluxes, and the QLP default detrended KSPSAP flux (now called DET\_FLUX) and the SPOC pre-search data conditioning PDCSAP fluxes. The KSPSAP and PDCSAP fluxes are detrended or filtered in a way that is intended to aid the search for exoplanet transits, but this generally plays havoc with eclipsing binary, and some other light curves. Even the most cursory examination of the light curves shows that the QLP-KSPSAP fluxes are very different, with the dominant variation being largely filtered out. The other data sets show less obvious differences through the sectors, but there are gradients between different orbits, particularly for the SPOC-PDCSAP data, and even for the SAP fluxes. The light curves of the QLP-SAP fluxes are shown in Fig.~\ref{fig:epoch}.

The data have been processed in a variety of different ways in an effort to find the most consistent set of frequencies, given the complexity of both the variations and the data. In the analysis of the individual sectors, it was found that different gradients through the orbits generated significant low-frequency features that compromised the detection of other, apparently real, weak frequencies. To address this issue a filtered QLP-SAP data set was created to suppress the low-frequency variations by subtracting a boxcar average with a window of 2\,d, which progressively filters the frequencies at $f < 0.5$\,c/d. In all cases the apparent gradients through each sector were also removed. All three sectors were also analysed together, again using the filtered, detrended data. 

The different data sets have been analysed by repeated application of a period search -- fitting and subtraction loop down to some arbitrary limit. Frequencies were identified through both a Discrete Fourier Transform (DFT) and a Lomb-Scargle periodogram, and then refined and removed using a multi-frequency Fourier fitting routine of the form
\begin{equation}
	m_{k}=\sum_{j=1}^{n}  a_{j}\cos( 2\pi f_j T_k+\phi_{j}) + c_{z} 
\end{equation}
where $m_k$ is the observed magnitude at time $T_k$, $j=1,...,n$ is the number of frequencies, with just the single harmonic for each frequency, $\phi_{j}$ is the phase offset for each frequency $j$, and $c_{z}$ is the constant level for any particular subset of the data. Each sector was first treated separately, and the frequency search was continued until the features were only occurring in one data set. However, some of these single-sector frequencies did appear before some of the multi-sector features. Most of the frequencies identified were common between the raw QLP and SPOC SAP, and the filtered QLP-SAP data, but this not always the case. In general, the precision on the frequencies from the individual sectors is relatively poor, typically 0.004\,c/d, and is limited in part by the short runs of data. It also became clear that the noise level in sector 60, which contains most of the data, is significantly worse than the other two sectors.

\begin{figure}[!t]
	\centering
	\includegraphics[width=0.49\textwidth]{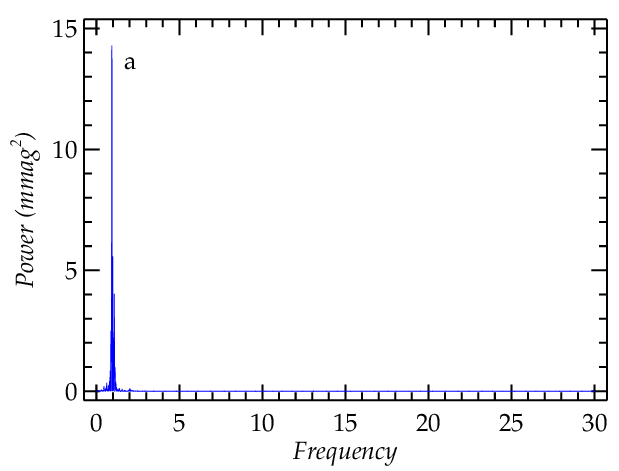}
	\includegraphics[width=0.49\textwidth]{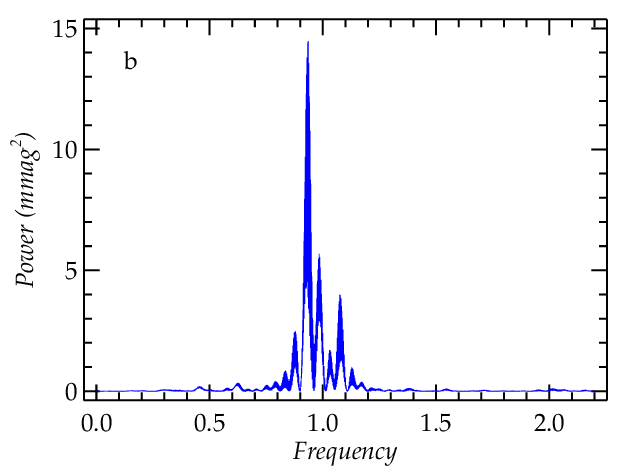}
	\includegraphics[width=0.49\textwidth]{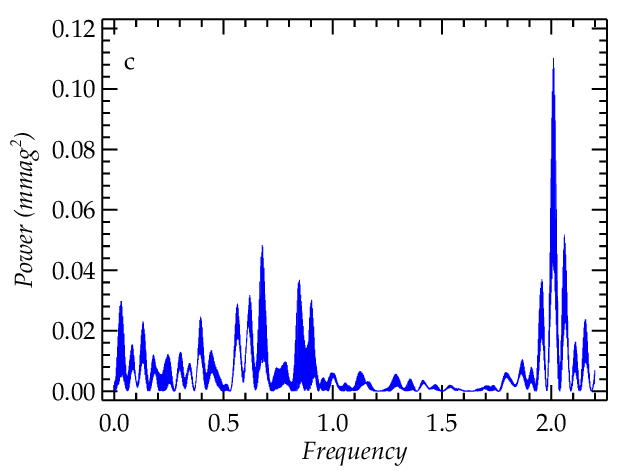}
	\includegraphics[width=0.49\textwidth]{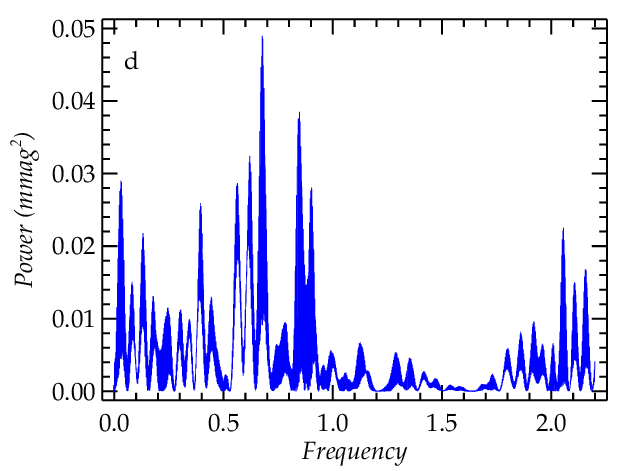}
	\caption{The DFT periodograms of the combined filtered QLP-SAP data. The four plots show the DFT of the data prior to the removal of any frequencies to 30\,c/d (a) and the expanded region to 2.2\,c/d (b), where all the power is concentrated, and (c) after the removal of the three frequencies near $f=1$\,c/d, so showing $f_4$, then (d) after four frequencies, showing $f_5$ and higher frequencies. The dramatic difference in the power of the main frequencies is clear, and also the potential confusion of the weakest features. \label{fig:dft}}
\end{figure}

The periodogram of the combined filtered data set (see Fig.~\ref{fig:dft})showed significant fringing around each feature due to the 1-yr alias (at 0.0027\,c/d) and this seriously hampered the identification of the 'correct' frequency. There was often little discrimination between the alternatives and the process became quite fragile. To resolve this issue, it was necessary to test the different frequency trees, but this was limited to for the stronger features. For the weaker features this process was not followed due to the myriad of possibilities, and the alias suggested by the periodogram was used. The danger of this approach is that false features can be generated if the incorrect alias is removed. The frequency search was continued until all the frequencies from multiple sectors were identified.

\section*{Results}

\begin{figure}[!b]
	\centering
	\includegraphics[width=0.49\textwidth]{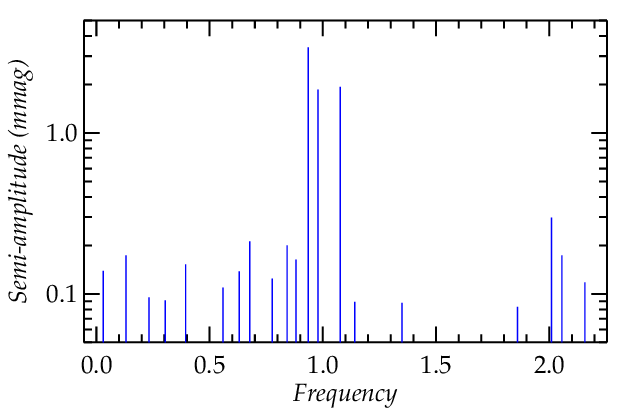}
	\includegraphics[width=0.49\textwidth]{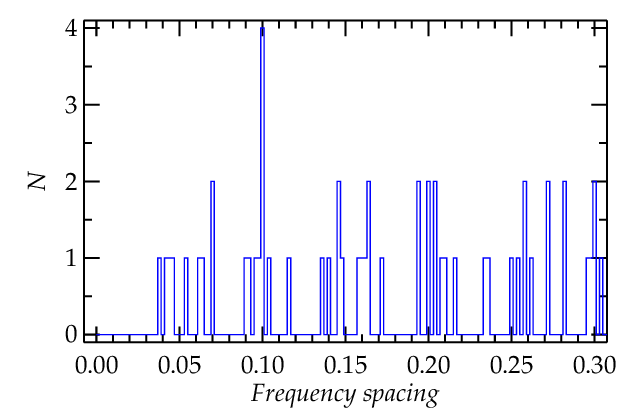}
	\caption{(Left) The frequency and amplitude distribution of the frequencies found in the combined filtered QLP-SAP data listed in Table~\ref{tab:freq}. The amplitude scale is logarithmic. (Right) The histogram of the frequency spacings.  \label{fig:freq}}
\end{figure}

The main arbiter of significance was the repeated detection of a particular frequency in the different sectors, but even this has its limitations due to the poor frequency resolution of each subset. Ultimately 19 frequencies have been identified and these are listed in Table~\ref{tab:freq}, but there is still some ambiguity. The table lists each frequency in order of detection in the combined QLP-SAP detrended filtered data, with the most likely corresponding frequencies found in the individual sectors. Some of these associations may seem a little loose. If a consistent frequency is present in the raw QLP-SAP data or SPOC-SAP data then that is indicated by * or \#. These data are not independent, but they do help to provide confirmation that the features found are not just a product of the filtering process. 

The main issue is the ambiguity in $f_3$, which in the alternate tree corresponds to the lower 1-yr alias of the frequency in the primary listing. Although the fits using this alias generate almost identical residuals, it is relegated to an alternative as the uncertainty on the frequency is about three times the optimal value. Perhaps surprisingly, most of the frequencies in this alternative tree are also present in the primary tree, except for the weak features, $f_{15}$, $f_{16}$ and $f_{18}$, suggesting they may be quite robust.

\begin{table}[!t]
	\caption{Table of frequencies recovered from the TESS QLP SAP filtered data}	
	\label{tab:freq}
	\centering
	\begin{tabular}{lllcllc} 
		\hline
Id  & Frequency($\sigma$) & A($\sigma$) mmag & Source & Frequency($\sigma$) & A($\sigma$) mmag &Sector \\ 
\hline \hline   
\multirow{3}{*}{$f_1$~~*\#}&  \multirow{3}{*}{0.93555(5)}  & \multirow{3}{*}{3.469(8)} &   \multirow{3}{*}{All}  &   0.9342(15) &  3.249(20) &  47 \\
&              &          &        &   0.9349(16) &  3.505(22) &  20 \\
&              &          &        &   0.9383(28) &  3.779(32) &  60 \\
\hline
\multirow{3}{*}{$f_2$~~*\#}&  \multirow{3}{*}{1.07733(5)}  & \multirow{3}{*}{1.918(8)} &   \multirow{3}{*}{All}  &   1.0758(13) &  1.966(18) &  20 \\
&              &          &        &   1.0765(10) &  1.917(11) &  47 \\
&              &          &        &   1.0780(13) &  1.796(12) &  60 \\
\hline
\multirow{3}{*}{$f_3$~~*\#}&  \multirow{3}{*}{0.97862(6)}  & \multirow{3}{*}{1.848(8)} &   \multirow{3}{*}{All}  &   0.9731(36) &  2.101(34) &  60 \\
&              &          &        &   0.9749(22) &  1.936(20) &  20 \\
&              &          &        &   0.9763(22) &  1.916(14) &  47 \\
($f_3$)   &  0.97591(17) & 1.886(10) &   All  &    &    &    \\
\hline
\multirow{3}{*}{$f_4$~~*\#}&  \multirow{3}{*}{2.01018(5)}  & \multirow{3}{*}{0.339(8)} &   \multirow{3}{*}{All}  &   2.0082(26) &  0.267(11) &  47 \\
&              &          &        &   2.0085(20) &  0.404(10) &  60 \\
&              &          &        &   2.0161(111)&  0.270(43) &  20 \\
\hline
\multirow{2}{*}{$f_5$~~*\#}&  \multirow{2}{*}{0.67695(12)} & \multirow{2}{*}{0.219(8)} &   \multirow{2}{*}{All}  &   0.6641(95) &  0.254(19) &  20 \\
&              &          &        &   0.6843(40) &  0.212(12) &  47 \\
\hline
\multirow{2}{*}{$f_6$~~*\#}&  \multirow{2}{*}{0.84226(15)} & \multirow{2}{*}{0.202(8)} &   \multirow{2}{*}{All}  &   0.8312(40) &  0.379(13) &  47 \\
&              &          &        &   0.8417(47) &  0.178(13) &  60 \\
\hline
\multirow{2}{*}{$f_7$~~*} &  \multirow{2}{*}{0.03044(31)} & \multirow{2}{*}{0.130(7)} &   \multirow{2}{*}{All}  &   0.0376(56) &  0.155(15) &  60 \\
&              &          &        &   0.0376(62) &  0.117(15) &  20 \\
\hline
\multirow{2}{*}{$f_8$}   &  \multirow{2}{*}{0.39411(12)} & \multirow{2}{*}{0.156(8)} &   \multirow{2}{*}{All}  &   0.3951(23) &  0.242(10) &  60 \\
&              &          &        &   0.4094(33) &  0.086(10) &  47 \\
\hline
\multirow{2}{*}{$f_9$~~*\#} & \multirow{2}{*}{2.05603(21)}  & \multirow{2}{*}{0.169(7)}  &   \multirow{2}{*}{All} &   2.0427(141)&  0.162(38) &  20 \\
&              &          &        &   2.0520(28) &  0.236(10) &  47 \\
\hline
\multirow{3}{*}{$f_{10}$~*} &  \multirow{3}{*}{0.13120(17)}  & \multirow{3}{*}{0.173(7)} &  \multirow{3}{*}{All} &   0.1328(41) &  0.206(12) &  60 \\
&              &          &        &   0.1371(28) &  0.088(10) &  47 \\
&              &          &        &   0.1464(9)  &  0.207(19) &  20 \\
\hline
$f_{11}$   &  0.55932(15) & 0.134(7) &   All  &    &    &    \\
\hline
$f_{12}$~*\# & 2.15754(20)   & 0.129(7)&   All & 2.1517(20)   & 0.144(10) &  47 \\
\hline
\multirow{2}{*}{$f_{13}$~*} &  \multirow{2}{*}{0.30389(13)}   &  \multirow{2}{*}{0.128(7)} &   \multirow{2}{*}{All} &   0.3043(24) &  0.159(10) &  60 \\
&              &          &        &   0.3057(48) &  0.210(18) &  20 \\
\hline
\multirow{2}{*}{$f_{14}$~*\#}&  \multirow{2}{*}{0.88158(31)}  & \multirow{2}{*}{0.131(8)} &   \multirow{2}{*}{All} &   0.8701(37) &  0.318(18) &  20 \\
&              &          &        &   0.8795(35) &  0.356(16) &  47 \\
\hline
\multirow{3}{*}{$f_{15}$~*} &  \multirow{3}{*}{0.63092(32)}  & \multirow{3}{*}{0.142(8)} &   \multirow{3}{*}{All} &   0.6181(21) &  0.257(12) &  60 \\
&              &          &        &   0.6314(121)&  0.104(24) &  20 \\
&              &          &        &   0.6409(38) &  0.147(11) &  47 \\
\hline
$f_{16}$   &  0.77700(34) & 0.125(8) &   All  &    &    &    \\
\hline
$f_{17}$   &  0.23244(29) & 0.095(7) &   All  &    &    &    \\
\hline
$f_{18}$~*\# &      1.34973(19)     &   0.088(7)     &   All &  1.3563(28)  & 0.091(10) &  47 \\
\hline
~~~~~*\# &           &        &   solo &  1.9226(27)  & 0.133(10) &  60 \\
\hline
\end{tabular}
%
\vspace{2mm}

* frequency present in original unfiltered QLP SAP data \\
\# frequency present in original unfiltered SPOC SAP data

\end{table}

The first ten frequencies each have multiple detections in the individual sectors, and in the unfiltered data as well, but $f_{11}$ is the first that shows no previous appearances. Two other weak signals, $f_{16}$ and $f_{17}$, also do not appear in the individual sectors, and the last one in the table, which only appears in one sector, is not recovered from the combined data. These have low confidence. Given the issues with the gradients through the data and the filtering process, there has to be some uncertainty over the low-frequency features, but having said that, they have survived the multiple-appearance test, so they cannot be dismissed out of hand. The maximum semi-amplitude of any features beyond 2.2\,c/d up to 30\,c/d is less than 0.03 mmag. 

The frequencies listed in Table~\ref{tab:freq} are plotted with their semi-amplitudes in Fig.~\ref{fig:freq}. The three main features lie close to 1\,c/d, with $f_4$ and some weaker frequencies near 2\,c/d. There is also a relatively even spread of weak features between zero and 1\,c/d. The distribution does not appear to be totally random, with a similar spacing between $f_2$ and $f_3$, the two features near $f_4$ at $f=2.05$ and 2.15\,c/d, and the two lowest frequencies at $f=0.030$ and 0.131\,c/d. To search the spacings more systematically, the differences between all the frequencies up to 0.3\,c/d have been calculated and are plotted in Fig.~\ref{fig:freq}. These show a distinct concentration near 0.101\,c/d, but the statistics are poor, and this spacing does not appear to apply to many frequencies.

The frequencies and amplitudes of the variations place the star firmly in the gamma Doradus range, and more specifically in the SYM or ASYM classification of \citet{2011MNRAS.415.3531B}. The lack of any significant variation at higher frequencies means that this is not a delta Scuti star or a hybrid variable. See \cite{2025ApJ...978...53W} for a helpful comparison. The SYM and ASYN stars tend to have one or two dominant, closely-spaced frequencies, and the SYM stars appear to have, subjectively, more symmetrical light curves. They are the most common $\gamma$~Dor type, tend to be marginally cooler than the ASYM type, and have relatively low amplitudes compared to other variants. Compare Fig.~\ref{fig:epoch} and \ref{fig:dft} with Figs.~1 and 2 from \citet{2011MNRAS.415.3531B}. In the previous paper, \citet{2020OEJV..207....1L} examined the spectral energy distribution of NSV~4031 and concluded that it was best matched to a spectral type of F5\,IV, with little reddening, at a temperature $\teff \simeq 6500$\,K. The absolute magnitude was calculated as $\mv \simeq 2.4$, about a magnitude above the ZAMS, and consistent with the slightly evolved spectral type. More recent data from Gaia DR3 \citep{2023A&A...674A...1G} gives similar results with $\teff = 6734$\,K and $\mg = 2.2$, equivalent to $\mv = 2.3$. Gaia's extended Apsis processing chain \citep{2023A&A...674A..26C} gives $\log{L}/{\lsun} = 0.87$, which with $\log{\teff}=3.83$ places the star solidly within the cooler tail of the SYM sample in \citet{2011MNRAS.415.3531B}'s Fig.~4.
 
In short-period pulsating stars there is a complex relationship between pulsation and rotation, and the rotational properties of these variables are not the same as normal stars \citep[see \eg][]{2025ApJ...978...53W}. Rotation also has an effect on the appearance of the frequencies and can produce splitting in different forms. In $\gamma$~Dor stars the rotational period is typically close to the pulsation period, and this can lead to variety of frequency-spacing patterns \citep[see \eg][for some background and examples]{2015ApJS..218...27V}. While there is some suggestion of a pattern in Fig.~\ref{fig:freq} it is hardly convincing, but it would not be unexpected, and may become clearer with further observations.

\section*{Conclusion}

New TESS observations have shown that NSV 4031 is a low-amplitude $\gamma$~Dor variable with dominant clusters of frequencies near 1 and 2 c/d, and a spread of very weak features at lower frequencies. Its physical properties from Gaia confirm that is lies at the heart of the $\gamma$~Dor distribution, close to the red edge of the instability strip and above the ZAMS.

\section*{Acknowledgements}

The author is indebted to John Greaves for pointing out the complex variations in the TESS data.
This paper includes data collected by the TESS mission, which are publicly available from the Mikulski Archive for Space Telescopes (MAST). Funding for the TESS mission is provided by NASA’s Science Mission directorate.
The authors are pleased to acknowledge use of NASA's Astrophysics Data System Bibliographic Services.
This research has made use of the SIMBAD database and the VizieR catalogue access tool, CDS, Strasbourg, France (DOI: 10.26093/cds/vizier).

%
%

\end{document}